\documentclass[preprint,aps,prd,nofootinbib]{revtex4}
\usepackage{mathrsfs}

\usepackage{amsmath}
\usepackage{graphicx}
\usepackage{subfigure}
\usepackage{pifont}
\usepackage{txfonts}
\newcommand{\bea}{\begin{eqnarray}}
\newcommand{\eea}{\end{eqnarray}}
\newcommand{\beq}{\begin{equation}}
\newcommand{\eeq}{\end{equation}}

\def\k{{\vec k}}

\def\/{\over}

\begin{document}

\title{ Relationship between Hawking radiation \\ from black holes and spontaneous excitation of atoms}

\author{ Hongwei Yu and Wenting Zhou }
\affiliation{Department of Physics and Institute of  Physics,\\
Hunan Normal University, Changsha, Hunan 410081, China}

\begin{abstract}
Using the formalism that separates the contributions of vacuum
fluctuations and radiation reaction to the rate of change of the
mean atomic energy, we show that a two-level atom in interaction
with a quantum massless scalar field both in the Hartle-Hawking and
Unruh vacuum in a 1+1 dimensional black hole background
spontaneously excites as if there is thermal radiation at the
Hawking temperature emanating from the
 black hole. Our calculation, therefore, ties the existence of Hawking
radiation to the spontaneous excitation of a two-level atom placed
in vacuum in the exterior of a black hole and shows pleasing
consistence of two different physical phenomena, the Hawking
radiation and the spontaneous excitation of atoms, which are quite
prominent in their own right.

\end{abstract}

\maketitle


\paragraph{Introduction.}  Hawking
radiation from black holes, as one of the most striking effects that
arise from the combination of quantum theory and general relativity,
has attracted widespread interest in physics community. Currently,
several derivations of Hawking radiation have been proposed,
including Hawking's original one which calculates the Bogoliubov
coefficients between the quantum scalar field modes of the in vacuum
states  and those of the out vacuum~\cite{Hawking:rv,Hawking:sw}, an
elegant one based upon the Euclidean quantum
gravity~\cite{Gibbons:1976ue} which has been interpreted as a
calculation of tunneling through classically forbidden
trajectory~\cite{Parikh:1999mf}, the approach based upon string
theory~\cite{SC,Peet}, and a recent proposal which ties its
existence to the cancellation of gravitational anomalies at the
horizon~\cite{Robinson:2005pd}. Here we discuss yet another
 approach which calculates the spontaneous excitation rate of a
two-level atom interacting with massless quantum scalar fields in
vacuum states in a black hole background. Our investigation ties the
existence of Hawking radiation to the spontaneous excitation of a
two-level atom placed in vacuum in the exterior of a black hole,
thus revealing an interesting relationship between the existence of
Hawking radiation from black holes and the spontaneous excitation of
 atoms in vacuum.

Spontaneous emission, on the other hand, is one of the most
important features of atoms
 and so far mechanisms such as vacuum
fluctuations \cite{Welton48, CPP83}, radiation reaction
\cite{Ackerhalt73}, or a combination of them \cite{Milonni88} have
been put forward to explain why spontaneous emission occurs. The
ambiguity in physical interpretation arises because of the freedom
in the choice of ordering of commuting operators of the atom and
field in a Heisenberg picture approach to the problem. The
controversy was resolved when Dalibard, Dupont-Roc and
Cohen-Tannoudji(DDC) \cite{Dalibard82,Dalibard84} proposed a
formalism which distinctively separates the contributions of vacuum
fluctuations and radiation reaction by demanding a symmetric
operator ordering of atom and field variables.  The DDC formalism
has recently been generalized to study the spontaneous excitation of
uniformly accelerated atoms in interaction with vacuum scalar and
electromagnetic fields in a flat spacetime~\cite{Audretsch94,H. Yu,
ZYL06}, and these studies show that when an atom is accelerated, the
delicate balance between vacuum fluctuations and radiation reaction
that ensures the ground state atom's stability in vacuum is altered,
making possible the transitions to excited states for ground-state
atoms even in vacuum. In this paper, we apply this generalized DDC
formalism to investigate the spontaneous excitation of an atom held
static in the exterior region of a black hole and interacting with
vacuum quantum massless scalar fields in two dimensions,  and show
that the atom spontaneously excites as if it were irradiated  by or
immersed in a thermal radiation at the Hawking temperature,
depending on whether  the scalar field is in the Unruh or the
Hartle-Hawking vacuum. In other words, atoms feel the Hawking
radiation from black holes

\paragraph{Formalism.} When vacuum fluctuations are concerned in a curved spacetime, a
delicate issue then arises as to how the vacuum state of the
massless scalar field is determined. Normally, the vacuum state is
associated with non-occupation of positive frequency modes. However,
the positive frequency of field modes are defined with respect to
the time coordinate. Therefore, to define positive frequency, one
has to first specify a definition of time. In a spherically
symmetric black hole background, one definition is the Schwarzschild
time  and it is a natural definition of time in the exterior region.
However, the vacuum state associated with this choice of time
coordinate (Boulware vacuum) becomes problematic in the sense that
the expectation value of the energy-momentum tensor, evaluated in a
free falling frame, diverges at the horizon. Other possibilities
that avoid this problem are the Unruh vacuum~\cite{Unruh} and the
Hartle-Hawking vacuum~\cite{Hartle-Hawking}. The Unruh vacuum is
defined by taking modes that are incoming from $\mathscr{J}^-$ to be
positive frequency with respect to the Schwarzschild time, while
those that emanate from the past horizon are taken to be positive
frequency with respect to the Kruskal coordinate $\bar u$, the
canonical affine parameter on the past horizon. The Unruh vacuum is
regarded as the vacuum state that best approximates the state that
would obtain following the gravitational collapse of a massive body.
The Hartle-Hawking vacuum, on the other hand, is defined by taking
the incoming modes to be positive frequency with respect to $\bar
v$, the canonical affine parameter on the future horizon, and
outgoing modes to be positive frequency with respect to $\bar u$.
Let us note that the Hartle-Hawking state does not correspond to our
usual notion of a vacuum since it has thermal radiation incoming to
the black hole from infinity and describes a black hole in
equilibrium with a sea of thermal radiation.

Consider, in two dimensions,  a two-level atom in interaction with a
quantum real massless scalar field in a spherically symmetric black
hole background, of which the metric is given by
\begin{equation}
ds^2= \bigg(1-{2M\over r}\bigg)\;du\,dv={2M\over r}e^{-r/2M}d\bar
u\,d\bar v\;,
\end{equation}
where
 \beq
 u=t-r^*,\;\; v=t+r^*,\;\; r^* = r+ 2M\ln [(r/2M)-1],\; \; \bar u=-e^{-\kappa u}/\kappa, \;\;\bar v=e^{-\kappa
 v}/\kappa\;.
 \eeq
 Here $\kappa = 1/4M$ is the surface gravity of the black hole.
Without loss of generality, let us assume a pointlike two-level atom
on a stationary space-time trajectory $x(\tau)$, where $\tau$
denotes the proper time on the trajectory. The stationarity of the
trajectory guarantees the existence of stationary atomic states, $|+
\rangle$ and $|- \rangle$, with energies $\pm{1\/2}\omega_0$ and a
level spacing $\omega_0$.  The atom's Hamiltonian which controls the
time evolution with respect to $\tau$ is given, in Dicke's notation
\cite{Dicke54},  by
 \beq H_A (\tau)
=\omega_0 R_3 (\tau)\;,  \label{atom's Hamiltonian}
 \eeq
 where $R_3 =
{1\/2} |+ \rangle \langle + | - {1\/2}| - \rangle \langle - |$ is
the pseudospin operator commonly used in the description of
two-level atoms\cite{Dicke54}. The free Hamiltonian of the quantum
 scalar field that governs its time evolution with respect to $\tau$
is
 \beq
  H_F (\tau) = \int
d^3 k\, \omega_\k \,a^\dagger_\k\, a_\k\,
    {dt\/d \tau}\;. \label{free Hamiltonian}
 \eeq
Here $a^\dagger_\k$, $a_\k$ are the creation and annihilation
operators with momentum $\k$. Following Ref. \cite{Audretsch94}, we
assume that the interaction between the atom and the quantum field
is described by a Hamiltonian
 \beq
 H_I (\tau) = c \,R_2 (\tau)\,\phi (
x(\tau)) = \mu\; \omega_0\, R_2 (\tau) \,\phi ( x(\tau))\;,
\label{interaction Hamiltonian}
 \eeq
 where $c$ is a
coupling constant which we assume to be small, $R_2 = {1\/2} i ( R_-
- R_+)$, and $R_+ = |+ \rangle \langle - |$, $R_- = |- \rangle
\langle +|$. The coupling is effective only on the trajectory
$x(\tau)$ of the atom. Note that here we have defined a
dimensionless parameter, $\mu=c/\omega_o$.

We can now write down the Heisenberg equations of motion for the
atom and field observables. The field is always considered to be in
its vacuum state $|0 \rangle$.  We will separately discuss the two
physical mechanisms that contribute to the rate of change of atomic
observables: the contribution of vacuum fluctuations and that of
radiation reaction. For this purpose, we can split the solution of
field $\phi$ of the Heisenberg equations  into two parts:  a free or
vacuum part $\phi^f$, which is present even in the absence of
coupling, and a source part $\phi^s$, which represents the field
generated by the interaction between the atom and the field.
Following DDC\cite{Dalibard82,Dalibard84}, we choose a symmetric
ordering between atom and field variables and consider the effects
of $\phi^f$ and $\phi^s$ separately in the Heisenberg equations of
an arbitrary atomic observable G. Then, we obtain the individual
contributions of vacuum fluctuations and radiation reaction to the
rate of change of G. Since we are interested in the spontaneous
excitation of the atom,  we will concentrate on the mean atomic
excitation energy $\langle H_A(\tau) \rangle$. The contributions of
vacuum fluctuations(vf) and radiation reaction(rr) to the rate of
change of $\langle H_A \rangle$ can be written as ( cf.
Ref.\cite{Dalibard82,Dalibard84,Audretsch94,H. Yu,ZYL06} )
 \bea
 \left\langle {d H_A (\tau) \/ d\tau}
\right\rangle_{vf} &=&
    2 i\, c^2  \int_{\tau_0}^\tau d \tau' \, C^F(x(\tau),x(\tau'))
    {d\/ d \tau} \chi^A(\tau,\tau')\;, \label{general form of vf}\\
\left\langle {d H_A (\tau) \/ d\tau} \right\rangle_{rr} &=&
    2 i\, c^2
    \int_{\tau_0}^\tau d \tau' \, \chi^F(x(\tau),x(\tau')) {d\/
d \tau}
    C^A(\tau,\tau')\;,
    \label{general form of rr}
 \eea
with $| \rangle = |a,0 \rangle$ representing the atom in the state
$|a\rangle$ and the field in the  vacuum state $|0 \rangle$. Here
the statistical functions of the atom, $C^{A}(\tau,\tau')$ and
$\chi^A(\tau,\tau')$, are defined as \bea C^{A}(\tau,\tau') &=&
{1\/2} \langle a| \{ R_2^f (\tau), R_2^f (\tau')\}
    | a \rangle\;,\label{general form of Ca} \\
\chi^A(\tau,\tau') &=& {1\/2} \langle a| [ R_2^f (\tau), R_2^f
(\tau')]
    | a \rangle \;,\label{general form of Xa}
 \eea
 and those of the field are as
 \bea
 C^{F}(x(\tau),x(\tau')) &=& {1\/2}{\langle} 0| \{ \phi^f
(x(\tau)), \phi^f(x(\tau')) \} | 0 \rangle\;, \label{general form of
Cf}\\
\chi^F(x(\tau),x(\tau')) &=& {1\/2}{\langle} 0| [
\phi^f(x(\tau)),\phi^f (x(\tau'))] | 0 \rangle\;. \label{general
form of Xf}
 \eea
$C^A$ is called the symmetric correlation function of the atom in
the state $|a\rangle$, $\chi^A$ its linear susceptibility.  $C^F$
and $\chi^F$ are the Hadamard function and Pauli-Jordan or
 Schwinger function of the field respectively.
The explicit forms of the statistical functions of the atom are
given by \bea C^{A}(\tau,\tau')&=&{1\/2} \sum_b|\langle a | R_2^f
(0) | b
    \rangle |^2 \left( e^{i \omega_{ab}(\tau - \tau')} + e^{-i
\omega_{ab}
    (\tau - \tau')} \right)\;, \label{explicit form of Ca}\\
\chi^A(\tau,\tau') & =& {1\/2}\sum_b |\langle a | R_2^f (0) | b
\rangle |^2
    \left(e^{i \omega_{ab}(\tau - \tau')} - e^{-i \omega_{ab}(\tau -
\tau')}
    \right)\;, \label{explicit form of Xa}\eea
where $\omega_{ab}= \omega_a-\omega_b$ and the sum runs over a
complete set of atomic states.

\paragraph{Spontaneous excitation of atoms.} First let us apply the above formalism to the case of the Hartle-Hawking vacuum
for the scalar field.  Consider an atom held static at a radial
distance $R$ from the black hole. The Wightman function for massless
scalar fields in the Hartle-Hawking vacuum  in (1+1) dimension is
given by~\cite{QFT}
\begin{eqnarray}
D_H^{+}\,(x,x')=-\frac{1}{4\pi}\ln{\frac{4e^{2\kappa
R^*}\sinh^2\biggl(\kappa_R\;(\Delta
   \tau/2)-i\varepsilon\biggr)}{\kappa^2}}\;,
\end{eqnarray}
where
\begin{equation}
\Delta\tau=\Delta\,t \sqrt{g_{00}}=\Delta\,t
\sqrt{1-\frac{2M}{R}}\;,\;\;
\kappa_R=\frac{\kappa}{\sqrt{1-\frac{2M}{R}}}\;.
 \end{equation}
 This leads to the
following statistical functions of the scalar field
\begin{eqnarray}
C^F(x\,(\tau),x\,(\tau')\,) =-\frac{1}{8\pi}\,
   \biggl[2\ln\frac{4e^{2\kappa R^*}}{\kappa^4}+
   \ln\sinh^2\biggl(\frac{\kappa_R}{2}\Delta\tau-i\varepsilon\biggr)
   +\ln\sinh^2\biggl(\frac{\kappa_R}{2}\Delta\tau+i\varepsilon\biggr)\biggr]\;,
\end{eqnarray}
\begin{eqnarray}
\chi^F(x\,(\tau),x\,(\tau')) =\,-\frac{1}{8\pi}\,
   \biggl[\,\ln\sinh^2\biggl(\frac{\kappa_R}{2}\Delta\tau-i\varepsilon\biggr)
   -\ln\sinh^2\biggl(\frac{\kappa_R}{2}\Delta\tau+i\varepsilon\biggr)\biggr]\;.
\end{eqnarray}
Plugging the above expressions into Eq.~(\ref{general form of vf})
and Eq.~(\ref{general form of rr}) and performing the integration
yields the contribution of the vacuum fluctuations to the rate of
change of the mean atomic energy for the atom held static at a
distance $R$ from the black hole
\begin{eqnarray}
\left\langle\;\frac{dH_A(\tau)}{d\tau}\right\rangle_{vf}\,&=&-\mu^2\,\biggl[\sum_{\omega_a>\omega_b}\;\omega_0^2\;|\langle
a|R_2^f(0)|b\rangle|^2\;
    \biggl(\,\frac{1}{2}+\frac{1}{e^{\,\omega_{ab}\,(2\pi/\kappa_R)}-1}\biggr)\nonumber\\
    &&\quad\quad-\sum_{\omega_a<\omega_b}\;\omega_0^2\;|\langle
    a|R_2^f(0)|b\rangle|^2\;
    \biggl(\,\frac{1}{2}+\frac{1}{e^{\,|\,\omega_{ab}|\,(2\pi/\kappa_R)}-1}\biggr)\;\biggr]\;,
\end{eqnarray}
and that of radiation reaction
\begin{eqnarray}
\left\langle\frac{dH_A(\tau)}{d\tau}\right\rangle_{rr}\,=-\mu^2\,\biggl[\sum_{\omega_a>\omega_b}\frac{\omega_0^2}{2}\,
   |\langle a|R_2^f(0)|b\rangle|^2+\sum_{\omega_a<\omega_b}\frac{\omega_0^2}{2}\,|\langle
   a|R_2^f(0)|b\rangle|^2\biggr]\;.
\end{eqnarray}
Here we have extended the integration range to infinity for
sufficiently long times $\tau-\tau_0$. Adding up two contributions,
we obtain the total rate of change of the mean atomic energy
\begin{eqnarray}
\left\langle\frac{dH_A(\tau)}{d\tau}\right\rangle_{tot}=-\mu^2\,\biggl[\sum_{\omega_a>\omega_b}\;\omega_0^2\;|\langle
a|R_2^f(0)|b\rangle|^2
   \,\biggl(\,1+\frac{1}{e^{\,\omega_{ab}\,(2\pi/\kappa_R)}-1}\biggr)\nonumber\\
   -\sum_{\omega_a<\omega_b}
   \;\omega_0^2\;|\langle
   a|R_2^f(0)|b\rangle|^2\;\frac{1}{e^{\,|\,\omega_{ab}|\,(2\pi/\kappa_R)}-1}\,\biggr]\;.\label{totH}
\end{eqnarray}
One can see that, for a ground state atom held static at a radial
distance $R$ from the black hole, only the second term (
$\omega_{a}< \omega_{b}$) contributes and this contribution is
positive, revealing that the atom spontaneously excites and thus
transitions from ground state to the excited states occur
spontaneously in the exterior region of the black hole. The most
striking feature is that the spontaneous excitation rate is what one
would obtain if the atom were immersed in a thermal bath of
radiation at the temperature
\begin{eqnarray}
T=\frac{\kappa}{2\pi}\frac{1}{\sqrt{1-\frac{2M}{R}}}=(g_{00})^{-1/2}\,T_{H}
    \;,
\end{eqnarray}
where $T_H=\kappa/2\pi$ is just the Hawking temperature of the black
hole. In fact, the above result is the well-known Tolman
relation~\cite{Tolman} which gives the proper temperature as
measured by a local observer. While for an atom at spatial infinity
($R\rightarrow \infty$), the temperature as felt by the atom, $T$,
approaches $T_H$. Therefore, an atom infinitely far away from the
black hole (in the asymptotic region) would spontaneously excite as
if in a thermal bath of radiation at the Hawking temperature.
However, as the atom approaches the horizon ($R\rightarrow 2M$), the
temperature $T$ diverges. This can be understood as the fact that
the atom must be in acceleration relative to the local free-falling
frame, which blows up at the horizon, to maintain at a fixed
distance from the black hole, and this acceleration gives rises to
additional thermal effect~\cite{Unruh}.

Now let us briefly discuss what happens if the Hartle-Hawking vacuum
is replaced by the Unruh vacuum. Then it is easy to show that the
statistical functions of the scalar field become
\begin{eqnarray}
C^F\,(x(\tau),x(\tau')\,)&=&-\,\frac{1}{8\pi}\,\biggl\{\,2\ln\frac{2e^{KR^*}}{k\sqrt{1-
   \frac{2M}{R}}}+\ln\,\biggl[\sinh\,\biggl(\frac{a}{2}\Delta\tau-i\varepsilon\biggr)\,
   \biggl(\Delta\tau-i\varepsilon\biggr)\,\biggr]+\nonumber\\&&\ln\,\biggl[\sinh\,\biggl
   (\frac{a}{2}\Delta\tau+i\varepsilon\biggr)\,\biggl(\Delta\tau+i\varepsilon\biggr)
   \,\biggr]\,-a(\tau+\tau')\biggr\}\;,
\end{eqnarray}
\begin{eqnarray}
\chi^F\,(x(\tau),x(\tau')\,)=-\,\frac{1}{8\pi}\,\biggl\{\,\ln\,\biggl[\sinh\,\biggl
   (\frac{a}{2}\Delta\tau-i\varepsilon\biggr)\,\biggl(\Delta\tau-i\varepsilon\biggr)\,
   \biggr]-\ln\,\biggl[\sinh\,\biggl(\frac{a}{2}\Delta\tau+i\varepsilon\biggr)\,
   \biggl(\Delta\tau+i\varepsilon\biggr)\,\biggr]\,\biggr\}\;,
\end{eqnarray}
and the contribution of the vacuum fluctuations to the rate of
change of the mean atomic energy for the atom held static at a
distance $R$ from the black hole is given by
\begin{eqnarray}
\left\langle\frac{dH_A(\tau)}{d\tau}\right\rangle_{vf}\,&=&-\mu^2\,\biggl[\sum_{\omega_a>\omega_b}\;\omega_0^2\;
   |\langle a|R_2^f(0)|b\rangle|^2\,\biggl(\,{1\over 2}+{1\over 2}\,\frac{1}{e^{\,\omega_{ab}\,(2\pi/\kappa_R)}-1}\biggr)
   \nonumber\\
   &&\quad\quad-\sum_{\omega_a<\omega_b}\;\omega_0^2\;|\langle a|R_2^f(0)|b\rangle|^2\,
   \biggl(\,{1\over 2}+{1\over 2}\,\frac{1}{e^{\,|\,\omega_{ab}|\,(2\pi/\kappa_R)}-1}\biggr)\,\biggr]\;,
\end{eqnarray}
and that of radiation reaction by
\begin{eqnarray}
\bigg\langle\frac{dH_A(\tau)}{d\tau}\bigg\rangle_{rr}\,=-\mu^2\,\biggl[\sum_{\omega_a>\omega_b}
   \, {\omega_0^2\over 2}\,|\langle a|R_2^f(0)|b\rangle|^2+\sum_{\omega_a<\omega_b}\,{\omega_0^2\over
    2}\,|\langle a|R_2^f(0)|b\rangle|^2\,\biggr]\;.
\end{eqnarray}
Consequently, we obtain, by adding up two contributions, the total
rate of change of the mean atomic energy
\begin{eqnarray}
\bigg\langle\frac{dH_A(\tau)}{d\tau}\bigg\rangle_{tot}&=&-\mu^2\,\biggl[\sum_{\omega_a>\omega_b}\,\omega_0^2\;
   |\langle a|R_2^f(0)|b\rangle|^2\,\biggl(\,1+{1\over 2}\,\frac{1}{e^{\,\omega_{ab}\,(2\pi/\kappa_R)}-1}\biggr)
   \nonumber\\
   &&\quad\quad-\sum_{\omega_a<\omega_b}\;\omega_0^2\,|\langle a|R_2^f(0)|b\rangle|^2\,{1\over 2}\,
   \frac{1}{e^{\,|\,\omega_{ab}|\,(2\pi/\kappa_R)}-1}\,\biggr]\;.
\end{eqnarray}
It is interesting to note that the term of the thermal like
contribution is half of that in the Hartle-Hawking case (refer to
Eq.~(\ref{totH})). This is consistent with our understanding that
the Unruh vacuum corresponds to the state following the collapsing
of a massive body to form a black hole, and as a result, the atom,
held static at a radial distance $R$, spontaneously excites as if it
were irradiated by a beam of outgoing thermal radiation at the
temperature $T=\kappa_R/2\pi$, in other words, atoms feel the
Hawking radiation. While the Hartle-Hawking vacuum is the state that
includes a thermal radiation at the Hawking temperature {\it
incoming} from infinity and describes an eternal black hole in
thermal equilibrium with the incoming thermal radiation. Therefore
the spontaneous excitation rate doubles.

\paragraph{Summary}

By evaluating the rate of change of the mean atomic energy for a
two-level atom in interaction with a massless scalar field in both
the Hartle-Hawking and the Unruh vacuum in a 1+1 dimensional black
hole background, we have demonstrated that an inertial atom far away
from the black hole (in the asymptotic region) would spontaneously
excite as if there is thermal radiation at the Hawking temperature
emanating from the black hole, or in other words, atoms feel the
Hawking radiation from black holes.  Therefore, our discussion can
be considered as  providing another approach to the derivation of
the Hawking radiation. Our result also reveals an interesting
relationship between the existence of Hawking radiation from black
holes and the spontaneous excitation of a two-level atom in vacuum
in the exterior of a black hole, and shows pleasing consistence of
two different physical phenomena, the Hawking radiation and the
spontaneous excitation of atoms, which are quite prominent in their
own right.

This work was supported in part  by the National Natural Science
Foundation of China  under Grants No.10375023 and No.10575035, and
the Program for New Century Excellent talents in
University(No.04-0784)


\begin{thebibliography}{99}


\bibitem{Hawking:rv}
S.~Hawking,
Nature (London) {\bf 248}, 30 (1974).


\bibitem{Hawking:sw}
S.~Hawking,
Commun.\ Math.\ Phys.\  {\bf 43}, 199 (1975).

\bibitem{Gibbons:1976ue}
G.~Gibbons and S.~Hawking,
Phys.\ Rev.\ D {\bf 15}, 2752 (1977).

\bibitem{Parikh:1999mf}
M.~Parikh and F.~Wilczek,
Phys.\ Rev.\ Lett.\  {\bf 85}, 5042 (2000).

\bibitem{SC}
A.~Strominger and C.~Vafa, Phys. Lett. {\bf B 379}, 99(1996).

\bibitem{Peet}A.~Peet, hep-th/0008241.

\bibitem{Robinson:2005pd}
S.~P.~Robinson and F.~Wilczek,
Phys.\ Rev.\ Lett.\  {\bf 95}, 011303 (2005); S.~Iso, H.~Umetsu, and
F.~Wilczek, Phys.\ Rev.\ Lett.\  {\bf 96}, 151302 (2006); Phys.\
Rev.\ {\bf D74},  044017 (2006).

\bibitem{Welton48}T. A. Welton, Phys. Rev. {\bf 74}, 1157(1948).
\bibitem{CPP83}G. Compagno, R. Passante and F. Persico, Phys. Lett. A {\bf
98},253(1983).
\bibitem{Ackerhalt73}J. R. Ackerhalt, P. L. Knight and J. H. Eberly, Phys. Rev.
Lett. {\bf 30}, 456(1973).
\bibitem{Milonni88}P. W. Milonni, Phys. Scr. {\bf T21}, 102(1988); P. W.
Milonni and W. A. Smith, Phys. Rev. A {\bf 11}, 814(1975).
\bibitem{Dalibard82} J. Dalibard, J. Dupont-Roc and C. Cohen-Tannoudji, J.
Phys. (France){\bf43}, 1617(1982).
\bibitem{Dalibard84} J. Dalibard, J. Dupont-Roc and C. Cohen-Tannoudji, J.
Phys. (France){\bf45}, 637(1984).

\bibitem{Audretsch94}J. Audretsch and R. M\"uller, Phys. Rev. A {\bf 50},
1755(1994).
\bibitem{H. Yu} H. Yu and S. Lu, Phys. Rev. D {\bf72},
064022(2005).
\bibitem{ZYL06} Z. Zhu, H. Yu and S. Lu, Phys. Rev. D {\bf73},
107501(2006).

\bibitem{Unruh} W.G. Unruh, Phys. Rev. D {\bf 14}, 870(1976).
\bibitem{Hartle-Hawking}J.~Hartle and S.~Hawking, Phys. Rev. {\bf
D13}, 2188 (1976).
\bibitem{Dicke54}R. H. Dicke Phys. Rev. {\bf 93}, 99(1954).

\bibitem{QFT} N. D. Birrell and P. C. W. Davies, {\it
Quantum Field Theory in Curved Space} (Cambridge Univ. Press,
Cambridge, 1982).
\bibitem{Tolman}R.~Tolman, Phys. Rev. {\bf 35}, 904 (1930); R.~Tolman and P.~Ehrenfest, Phys. Rev. {\bf 36}, 1791 (1930).

\end{thebibliography}
\end{document}